\documentclass[preprint,amsmath,amssymb,11pt,english,aps,showpacs,showkeys]{revtex4}
\usepackage{graphicx}
\usepackage{graphics}
\usepackage{amsmath}
\usepackage{dcolumn}
\usepackage{amssymb}
\usepackage{bm}
\usepackage[latin1]{inputenc}

\begin{document}
\title{Persistent spin currents in an elastic Landau system}
\author{K. Bakke}
\email{kbakke@fisica.ufpb.br}
\affiliation{Departamento de F\'isica, Universidade Federal da Para\'iba, Caixa Postal 5008, 58051-970, Jo\~ao Pessoa, PB, Brazil.}  

\author{C. Furtado}
\email{furtado@fisica.ufpb.br} 
\affiliation{Departamento de F\'isica, Universidade Federal da Para\'iba, Caixa Postal 5008, 58051-970, Jo\~ao Pessoa, PB, Brazil.}

\begin{abstract}
We consider a neutral particle with permanent magnetic dipole moment in an elastic medium with the presence of a uniform distribution of screw dislocations interacting with a radial electric field. We show that the uniform distribution of dislocations plays the role of an effective uniform magnetic field, and obtain a spectrum of energy which depends on the Aharonov-Casher geometric phase [Y. Aharonov and A. Casher, Phys. Rev. Lett. {\bf53}, 319 (1984)]. Moreover, from the dependence of energy levels on the Aharonov-Casher geometric phase, we calculate the persistent spin currents in this elastic Landau system. 
\end{abstract}


\maketitle

\section{Introduction}

Studies of the arising of persistent currents have attracted a great deal of attention in recent decades \cite{new2}. Persistent currents have been studied for spinless quantum particles confined to a quantum ring \cite{by}, two-dimensional quantum rings and quantum dots \cite{tan2} due to the presence of the Aharonov-Bohm quantum flux. In the presence of a disclination, the confinement of a spinless quantum particle to a two-dimensional quantum dot has been discussed in \cite{fur4}, where it has been shown that the presence of a topological defect changes the periodicity of the persistent currents. Persistent currents have also been studied from the dependence of the energy levels on the Berry phase \cite{ring2}, the Aharonov-Anandan quantum phase \cite{ring3}, and the Aharonov-Casher geometric phase \cite{ring1,ring4}. In particular, the Aharonov-Casher geometric phase \cite{ac} arises from the interaction between the permanent magnetic dipole moment of a neutral particle with an external electric field, therefore, persistent currents associated with the dependence of the energy levels on the Aharonov-Casher geometric phase \cite{ac} are called persistent spin currents \cite{ring1}. 

In recent years, studies of torsion effects on quantum systems have shown the coupling between spin and torsion \cite{shap,shap2}, an analogue effect of the Aharonov-Casher effect \cite{bf2}, and mathematical models for implementing one-qubit quantum gates for neutral particles have been proposed in \cite{bf7}. In crystalline solids, a torsion field is described by the Volterra process \cite{kleinert} which consists in describing a topological defect in a solid by the process of ``cut'' and ``glue'' of an elastic medium. In the continuum picture of defects, a torsion field corresponds to the strain and stress induced by a defect in an elastic medium. Examples of torsion in an elastic medium described by the Volterra process are the screw dislocation and the edge dislocation \cite{moraesG2}. An interesting study based on the continuum picture of defects in solids was proposed by Katanaev and Volovich \cite{kat} by using the differential geometry. In this approach, a linear topological defect in a solid, such as a screw dislocation, is described by the spatial part of a line element which corresponds to one particular solution of Einstein's equations in general relativity \cite{moraesG2,tt7}. Thereby, based on the Katanaev-Volovich approach \cite{kat}, studies of the influence of torsion on quantum systems have been made in the motion of an electron in a crystal \cite{aur,new}, quantum scattering \cite{tt11}, Landau levels for a nonrelativistic scalar particle \cite{tt13} and a spin-half neutral particle \cite{bf5}, Berry's phase \cite{tt10} and a two-dimensional quantum ring \cite{ani}. It is worth mentioning other studies of torsion in the classical point of view such as geodesics around a dislocation \cite{moraesG} and torsion effects on electromagnetic fields \cite{moraesG3}.

In this paper, we consider a neutral particle with a permanent magnetic dipole moment in an elastic medium containing a uniform distribution of screw dislocations \cite{f4}. This neutral particle interacts with an electric field produced by a linear distribution of electric charges on the symmetry axis of one screw dislocation placed at the center of the area containing the distribution of screw dislocations. The origin of this electric field can be explained by the presence of a charged dislocation. During the formation of a topological defect in a crystal, chemical bonds are broke and reconstructed. This process of breaking and reconstructing chemical bonds can produce some bonds which remain broken at the core of the defect, which are called dangling bonds. This process is the main reason for the electrical activity of dislocations in semiconductors. The existence of dangling bonds along the defect produces an electric field, therefore the dislocation acquires an electric charge density, which in turn, produces a electric field in the surrounding medium. In general, this electric charge density is screened by a cylindrical space charge of opposite sign, which is formed by ionized donors or acceptors. In this paper, we consider an electric field generated by a linear distribution of electric charges on the axis of the dislocation. This configuration is known as the Read cylinder \citep{read}. Recently, the Read cylinder has been considered in the study of the quantum scattering of a neutral particle by a charged screw dislocation in Ref. \cite{carlos}. It worth mentioning that Figielski {\it et al} \cite{figie1,figie2} found the solid state analogue of the Aharonov-Bohm effect \cite{ab}. We show that bound states can be achieved in the Aharonov-Casher system \cite{ac} where the uniform distribution of dislocations plays the role of an effective uniform magnetic field, and the spectrum of energy depends on the Aharonov-Casher geometric phase \cite{ac}. In this way, from the dependence of energy levels on the Aharonov-Casher geometric phase \cite{ac}, we calculate the persistent spin currents \cite{ring1} in this elastic Landau system.

This paper is organized as follows: in section II, we study the behaviour of a neutral particle with a permanent magnetic dipole moment interacting with an external electric field in an elastic Landau system. We obtain the energy levels for bound states and the persistent spin currents which arise from the dependence of the spectrum of energy on the Aharonov-Casher geometric phase \cite{ac}; in section III, we consider the presence of a hard-wall confining potential in the elastic Landau system, and obtain the energy levels for bound states analogous to confining a neutral particle to a quantum dot with a hard-wall confining potential \cite{dot,bf20}. We also calculate the persistent spin currents that arises from the dependence of the spectrum of energy on the Aharonov-Casher geometric phase \cite{ac}; in section IV, we present our conclusions.

\section{the Aharonov-Casher effect in an elastic Landau system}

In this section, we wish to study the behaviour of a neutral particle with a permanent magnetic dipole moment in an elastic Landau system when the magnetic dipole moment of the neutral particle interacts with an external electric field. Our goal is to obtain the energy levels for bound states in an elastic medium with the presence of a uniform distribution of screw dislocations, and show the dependence of the energy levels on the Aharonov-Casher geometric phase \cite{ac} which gives rise to the arising of persistent currents \cite{by}. First of all, let us consider an elastic medium in the absence of topological defects. The quantum dynamics of the neutral particle is described by the the following Schr\"odinger-Pauli equation \cite{ac,anan,anan2}
\begin{eqnarray}
i\frac{\partial\psi}{\partial t}=\frac{1}{2m}\,\vec{\sigma}\cdot\left[\vec{p}-i\mu\vec{E}\right]\,\vec{\sigma}\cdot\left[\vec{p}+i\mu\vec{E}\right]\psi,
\label{1.1}
\end{eqnarray}
where we consider the units $\hbar=c=1$ from now on. We have in (\ref{1.1}) that $\mu$ corresponds to the permanent magnetic dipole moment of the neutral particle, $m$ is the mass of the neutral particle and the matrices $\sigma^{i}$ correspond to the Pauli matrices. The Pauli matrices satisfy the relation $\left(\sigma^{i}\,\sigma^{j}+\sigma^{j}\,\sigma^{i}\right)=2\,\eta^{ij}$, where $\eta^{ij}=\mathrm{diag}(+ + +)$. An interesting quantum effect arises from the interaction between the magnetic dipole moment of the neutral particle and an electric field yielding the appearance of a geometric phase in the wave function of the neutral particle. Now, we consider the Aharonov-Casher effect generated by a neutral particle in the presence of a charged screw dislocation. This system consists in considering the magnetic dipole moment of the neutral particle being aligned to the $z$-axis, and an external electric field produced by a linear distribution of electric charges along the dislocation located in the $z$-axis. The electric field produced by a charged screw dislocation \cite{read,carlos,figie1,figie2} is given by $\vec{E}=\frac{\lambda}{\rho}\,\hat{\rho}$ (where $\rho=\sqrt{x^{2}+y^{2}}$, $\lambda$ is the linear charge distribution along the $z$-axis, and $\hat{\rho}$ is a unit vector on the radial direction). Thereby, we have that the interaction between the magnetic dipole moment and the radial electric field produced by a charged screw dislocation gives rise to a geometric phase given by 
\begin{eqnarray}
\phi_{\mathrm{AC}}=\oint\left(\vec{\mu}\times\vec{E}\right)_{k}\,dx^{k}=\pm2\pi\,\mu\lambda.
\label{1.2}
\end{eqnarray}
The quantum effect related to the geometric phase (\ref{1.2}) is called the Aharonov-Casher effect \cite{ac} \footnote{Note, by restoring $\hbar$ and $c$, we have $\phi_{\mathrm{AC}}=\frac{1}{\hbar c}\oint\left(\vec{\mu}\times\vec{E}\right)_{k}\,dx^{k}=\pm2\pi\frac{\mu\lambda}{\hbar c}$.}.

Now, let us introduce the geometric description of an elastic medium with the presence of a uniform distribution of screw dislocations. The geometric description of defects in solids was proposed by Katanaev and Volovich \cite{kat}. In the Katanaev-Volovich approach \cite{kat}, linear topological defects are described by the spatial part of the line element of a general relativity scenario. For instance, disclinations in solids correspond to the spatial part of the line element of the cosmic string spacetime \cite{kat,moraesG2}. Recently, by using the geometric theory of defects in solids, one-qubits quantum gates have been proposed in \cite{bf7}. In this work, we consider a uniform distribution of parallel screw dislocations in an elastic medium which is described by the line element \cite{f4}
\begin{eqnarray}
ds^{2}=d\rho^{2}+\rho^{2}d\varphi^{2}+\left(dz+\Omega\rho^{2}d\varphi\right)^{2}
\label{1.3}
\end{eqnarray}
where $\Omega=b_{i}\,\frac{A}{2}$, with $A$ being the area density of dislocations and $b_ {i}$ is the Burgers vector. In recent years, the influence of the classical background described by (\ref{1.3}) on the quantum dynamics of a spinless particle has yielded bound states corresponding to the elastic Landau levels \cite{f4}. Moreover, bound states have been obtained by a spin-half neutral particle both under the influence of a constant force field \cite{b6} and in the presence of a repulsive Coulomb-like potential \cite{bm}, due to the presence of a uniform distribution of parallel screw dislocations. 
 
By observing the symmetry involving the classical background given in (\ref{1.3}), we can work the Schr\"odinger-Pauli equation (\ref{1.1}) by using the mathematical formulation of the spinor theory in curved spacetime \cite{weinberg}. Spinors in curved spacetime are defined locally and obey the local Lorentz transformations. In this way, the procedure of defining spinors in curved spacetime consists in building a local reference frame via a noncoordinate basis $\hat{\theta}^{a}=e^{a}_{\,\,\,\mu}\left(x\right)\,dx^{\mu}$, whose components $e^{a}_{\,\,\,\mu}\left(x\right)$ are called triads \cite{weinberg,naka,kleinert}. 
Triads satisfy the relation $g_{\mu\nu}=e^{i}_{\,\,\,\mu}\left(x\right)e^{j}_{\,\,\,\nu}\left(x\right)\,\eta_{ij}$, where $\eta_{ij}=\mathrm{diag}\left(+++\right)$. In this work, we denote the cylindrical coordinates (related to the topological defect) by Greek indices, and the local reference frame of the observers (related to the flat space or absence of defects) by Latin indices. Furthermore, triads have an inverse given by $dx^{\mu}=e^{\mu}_{\,\,\,a}\left(x\right)\,\hat{\theta}^{a}$, where we have the relations:  $e^{a}_{\,\,\,\mu}\left(x\right)\,e^{\mu}_{\,\,\,b}\left(x\right)=\delta^{a}_{\,\,\,b}$ and $e^{\mu}_{\,\,\,a}\left(x\right)\,e^{a}_{\,\,\,\nu}\left(x\right)=\delta^{\mu}_{\,\,\,\nu}$. Here, we choose the following triad field:
\begin{eqnarray}
\hat{\theta}^{1}=d\rho;\,\,\,\hat{\theta}^{2}=\rho\, d\varphi;\,\,\,\hat{\theta}^{3}=dz+\Omega\rho^{2} d\varphi.
\label{1.4}
\end{eqnarray}

Returning to the geometric theory of defects in solids, screw dislocations are linear topological defects related to the presence of torsion in an elastic medium \cite{kleinert}. Due to the presence of torsion and curvilinear coordinates, the partial derivative $\partial_{\mu}$ given in the Schr\"odinger-Pauli equation (\ref{1.1}) becomes the covariant derivative of a spinor $\nabla_{\mu}=\partial_{\mu}+\Gamma_{\mu}\left(x\right)$ \cite{shap}, where $\Gamma_{\mu}\left(x\right)=\frac{i}{4}\,\left[\omega_{\mu ab}\left(x\right)+K_{\mu ab}\left(x\right)\right]\,\Sigma^{ab}$ is called the spinorial connection. For two-spinors, we have that $\Sigma^{ab}$ is defined in the form $\Sigma^{ab}=\frac{i}{2}\left[\sigma^{a},\sigma^{b}\right]$, where we define the matrix $\sigma^{0}=I$ being the $2\times2$ identity matrix and $\sigma^{i}$ being the usual Pauli matrices. Further, the connection 1-form $K_{\mu ab}\left(x\right)$ is related to the contortion tensor via the expression \cite{shap}:
\begin{eqnarray}
K_{\mu ab}\left(x\right)=K_{\beta\nu\mu}\left(x\right)\left[e^{\nu}_{\,\,\,a}\left(x\right)\,e^{\beta}_{\,\,\,b}\left(x\right)-e^{\nu}_{\,\,\,b}\left(x\right)\,e^{\beta}_{\,\,\,a}\left(x\right)\right].
\label{1.5}
\end{eqnarray}
Moreover, the contortion tensor is related with the torsion tensor via $K^{\beta}_{\,\,\,\nu\mu}=\frac{1}{2}\left(T^{\beta}_{\,\,\,\nu\mu}-T_{\nu\,\,\,\,\mu}^{\,\,\,\beta}-T^{\,\,\,\beta}_{\mu\,\,\,\,\nu}\right)$. Note that the torsion tensor is antisymmetric in the last two indices, while the contortion tensor is antisymmetric in the first two indices. Following these definitions, we can represent the torsion tensor in terms of three irreducible components \cite{shap}: the trace 4-vector $T_{\mu}=T^{\beta}_{\,\,\,\mu\beta}$, the axial 4-vector $S^{\alpha}=\epsilon^{\alpha\beta\nu\mu}\,T_{\beta\nu\mu}$, and the tensor $q_{\beta\nu\mu}$ which satisfy the conditions: $q^{\beta}_{\,\,\mu\beta}=0$ and $\epsilon^{\alpha\beta\nu\mu}\,q_{\beta\nu\mu}=0$. In this way, the torsion tensor becomes $T_{\beta\nu\mu}=\frac{1}{3}\left(T_{\nu}\,g_{\beta\mu}-T_{\mu}\,g_{\beta\nu}\right)-\frac{1}{6}\,\epsilon_{\beta\nu\mu\gamma}\,S^{\gamma}+q_{\beta\nu\mu}
$, and we can rewrite the connection 1-form (\ref{1.5}) in terms of these irreducible components \cite{shap}. We also have in the expression of the spinorial connection the presence of the connection 1-form $\omega_{\mu\,\,\,\,b}^{\,\,\,a}\left(x\right)$ which is, in general, related to the curvature in an elastic medium. Both connections 1-form $\omega_{\mu ab}\left(x\right)$ and $K_{\mu ab}\left(x\right)$ can be obtained by solving the Cartan structure equations $T^{a}=d\hat{\theta}^{a}+\omega^{a}_{\,\,\,b}\wedge\hat{\theta}^{b}$ \cite{naka}, where the operator $d$ corresponds to the exterior derivative, the symbol $\wedge$ means the wedge product, $\omega^{a}_{\,\,\,b}=\omega_{\mu\,\,\,\,b}^{\,\,\,a}\left(x\right)\,dx^{\mu}$, and $T^{a}=\frac{1}{2}\,T_{\,\,\mu\nu}^{a}\,dx^{\mu}\wedge dx^{\nu}$ is the torsion $2$-form. For instance, by solving the Cartan structure equations for the triads (\ref{1.4}), we obtain \cite{b6}
\begin{eqnarray}
\omega_{\varphi\,\,\,2}^{\,\,\,1}\left(x\right)&=&-\omega_{\varphi\,\,\,1}^{\,\,\,\,2}\left(x\right)=-1;\nonumber\\
[-2mm]\label{1.6}\\[-2mm]
T^{3}&=&2\Omega\rho\,\,d\rho\wedge d\varphi.\nonumber
\end{eqnarray}
From the component of the torsion $2$-form given in (\ref{1.6}), we obtain one non-null component of the axial $4$-vector $S^{0}=-4\Omega$ \cite{b6}. 

In this way, by changing the partial derivative for the covariant derivative of a spinor into the Dirac equation, and applying the Foldy-Wouthuysen approximation \cite{fw} in order to obtain the nonrelativistic limit of the Dirac equation, it has been shown in \cite{bf2} that the general expression for the Schr\"ondiger-Pauli equation (\ref{1.1}) involving torsion in curvilinear coordinates is given by ($\hbar=c=1$)
\begin{eqnarray}
i\frac{\partial\psi}{\partial t}&=&\frac{1}{2m}\left[\vec{p}+\vec{\Xi}\right]^{2}\psi-\frac{\mu^{2}E^{2}}{2m}\,\psi+\frac{\mu}{2m}\left(\vec{\nabla}\cdot\vec{E}\right)\psi\nonumber\\
[-2mm]\label{1.7}\\[-2mm]
&+&\mu\,\vec{\sigma}\cdot\vec{B}\,\psi+\frac{1}{8}\,\vec{\sigma}\cdot\vec{S}\,\psi,\nonumber
\end{eqnarray}
where the vector $\vec{\Xi}$ is defined in such a way that its components are given in the local reference frame (\ref{1.4}) by $\Xi_{k}=\mu\left(\vec{\sigma}\times\vec{E}\right)_{k}+\frac{1}{2}\,\sigma^{3}\,e^{\varphi}_{\,\,\,k}\left(x\right)-\frac{1}{8}\,S^{0}\,\sigma_{k}$. Note that the terms $\frac{1}{2}\,\sigma^{3}\,e^{\varphi}_{\,\,\,k}\left(x\right)$ and $-\frac{1}{8}\,S^{0}\,\sigma_{k}$ comes from the contribution of the spinorial connection, and $\vec{\Xi}$ plays the role of an effective potential vector. Moreover, in the absence of defects, we have that the term $\left(\vec{\sigma}\times\vec{E}\right)$ plays the role of the Aharonov-Casher potential vector \cite{ac}. As discussed in Refs. \cite{shap,shap2}, the components of the vector $\vec{\sigma}$ can be considered as internal degrees of freedom, that is, we can consider the components of $\vec{\sigma}$ corresponding to the spin of the neutral particle. In this way, we can see in Eq. (\ref{1.7}) that the axial 4-vector $S^{\mu}$ couples to spinors, while the trace 4-vector $T^{\mu}$ and the tensor $q_{\beta\nu\mu}$ decouple. The spin-torsion coupling is given by the last term of (\ref{1.7}).

In the Aharonov-Casher system \cite{ac}, the magnetic dipole moment of the neutral particle is considered to be aligned to the $z$-axis, then, by taking the electric field $\vec{E}=\frac{\lambda}{\rho}\,\hat{\rho}$ (produced by a linear distribution of electric charges on the symmetry axis of one screw dislocation, which can be considered to be placed at the center of the area $A$), we have from (\ref{1.2}) that $\mu\left(\vec{\sigma}\times\vec{E}\right)=\pm\frac{\phi_{\mathrm{AC}}}{2\pi}\,\hat{\varphi}$. Further, for the neutral particle moving in a region $\rho\neq0$, we have that the term $\vec{\nabla}\cdot\vec{E}$ of Eq. (\ref{1.7}) is null if we assume that the wave function of the neutral particle is well-behaved at the origin. In this way, the Schr\"odinger-Pauli equation (\ref{1.7}) becomes
\begin{eqnarray}
i\frac{\partial\psi}{\partial t}&=&-\frac{1}{2m}\left[\frac{\partial^{2}}{\partial\rho^{2}}+\frac{1}{\rho}\frac{\partial}{\partial\rho}+\frac{1}{\rho^{2}}\frac{\partial^{2}}{\partial\varphi^{2}}-2\Omega\,\frac{\partial^{2}}{\partial z\partial\varphi}\right]\psi\nonumber\\
&-&\frac{1}{2m}\,\left(1+\Omega^{2}\rho^{2}\right)\frac{\partial^{2}\psi}{\partial z^{2}}+\frac{1}{2m}\frac{i\sigma^{3}}{\rho^{2}}\frac{\partial\psi}{\partial\varphi}+\frac{1}{8m\rho^{2}}\,\psi\nonumber\\
[-2mm]\label{3.5}\\[-2mm]
&-&\frac{i\sigma^{3}}{m}\frac{\phi_{\mathrm{AC}}}{2\pi\rho^{2}}\,\frac{\partial\psi}{\partial\varphi}+\frac{i\sigma^{3}}{m}\,\frac{\phi_{\mathrm{AC}}}{2\pi}\Omega\frac{\partial\psi}{\partial z}-\frac{i\sigma^{3}}{2m}\,\Omega\frac{\partial\psi}{\partial z}\nonumber\\
&-&\frac{1}{2m}\frac{\phi_{\mathrm{AC}}}{2\pi\rho^{2}}\,\psi-\frac{i\sigma^{3}\Omega}{2m}\frac{\partial\psi}{\partial z}+\frac{\Omega^{2}}{8m}\,\psi+\frac{1}{2m}\left(\frac{\phi_{\mathrm{AC}}}{2\pi\rho}\right)^{2}\psi.\nonumber
\end{eqnarray}

We can see that $\psi$ is an eigenfunction of $\sigma^{3}$, whose eigenvalues are $s=\pm1$ and the Hamiltonian of the equation (\ref{3.5}) commutes with the operators $\hat{J}_{z}=-i\partial_{\varphi}$ \footnote{It has been shown in Ref. \cite{schu} that the $z$-component of the total angular momentum operator in cylindrical coordinates is given by $\hat{J}_{z}=-i\partial_{\varphi}$, whose eigenvalues are $j=l+\frac{1}{2}=\pm\frac{1}{2},\pm\frac{3}{2},\ldots$.} and $\hat{p}_{z}=-i\partial_{z}$. Therefore, we can write the solution of the Schr\"odinger-Pauli equation (\ref{3.5}) in terms of the eigenfunctions of the operators $\hat{J}_{z}$ and $\hat{p}_{z}$ \cite{schu}, that is, $\psi_{s}=e^{-i\mathcal{E}t}\,e^{i\left(l+\frac{1}{2}\right)\varphi}\,e^{ikz}\,R_{s}\left(\rho\right)$, where $l=0,\pm1,\pm2,\ldots$ and $k$ is a constant (we consider $k>0$ as discussed in \cite{bm}). Substituting this general solution into the second order differential equation (\ref{3.5}), we obtain
\begin{eqnarray}
\mathcal{E}R_{s}&=&-\frac{1}{2m}\left[\frac{d^{2}R_{s}}{d\rho^{2}}+\frac{1}{\rho}\frac{dR_{s}}{d\rho}\right]+\frac{1}{2m}\frac{\gamma_{s}^{2}}{\rho^{2}}R_{s}-\frac{\Omega k}{m}\gamma_{s}\,R_{s}\nonumber\\
[-2mm]\label{3.6}\\[-2mm]
&+&\frac{1}{2m}\left(k+s\frac{\Omega}{2}\right)^{2}\,R_{s}+\Omega^{2}k^{2}\rho^{2}\,R_{s},\nonumber
\end{eqnarray}
where we have defined the following parameter in the equation (\ref{3.6}): 
\begin{eqnarray}
\gamma_{s}=l+\frac{1}{2}\left(1-s\right)+s\frac{\phi_{\mathrm{AC}}}{\phi_{0}},
\label{3.6a}
\end{eqnarray}
where $\phi_{0}=2\pi$ \footnote{Note, by restoring $\hbar$ and $c$, we have $\phi_{0}=2\pi/\hbar\,c$ as in Ref. \cite{ring1}.}. Note that, the presence of a uniform distribution of screw dislocation yields no new term in the expression of the angular momentum (\ref{3.6a}) in contrast to the cases where there exist the presence of only one screw dislocation as pointed out in \cite{fm,az,f2,f3,ani} for a spinless quantum particle and in \cite{bf5} for a spin-half neutral particle. In the present case, the presence of the uniform distribution of screw dislocations provides the spin-torsion coupling as pointed out in \cite{shap}, and a term proportional to $\rho^{2}$ in Eq. (\ref{3.6}) which is analogous to a harmonic oscillator potential.

Next, we make the change of variables $\xi=\Omega k \rho^{2}$, then, the equation (\ref{3.6}) becomes
\begin{eqnarray}
\xi\,\frac{d^{2}R_{s}}{d\xi^{2}}+\frac{dR_{s}}{d\xi}-\frac{\gamma^{2}_{s}}{4\xi}\,R_{s}-\frac{\xi}{4}\,R_{s}+\frac{\beta_{s}}{4\Omega k}\,R_{s}=0,
\label{3.7}
\end{eqnarray}
where $\beta_{s}=2m\mathcal{E}+2\Omega k\gamma_{s}-\left(k+s\frac{\Omega}{2}\right)^{2}$. Now, we consider the wave function of the neutral particle is well-behaved at the origin, thus, the solution for the equation (\ref{3.7}) can be written as
\begin{eqnarray}
R_{s}\left(\xi\right)=e^{-\frac{\xi}{2}}\,\xi^{\frac{\left|\gamma_{s}\right|}{2}}\,M_{s}\left(\xi\right).
\label{3.8}
\end{eqnarray}
Substituting the solution (\ref{3.8}) into (\ref{3.7}), we obtain a second order differential equation given by
\begin{eqnarray}
\xi\,M_{s}''+\left[\left|\gamma_{s}\right|+1-\xi\right]\,M_{s}'+\left[\frac{\beta_{s}}{4\Omega k}-\frac{\left|\gamma_{s}\right|}{2}-\frac{1}{2}\right]\,M_{s}=0,
\label{3.9}
\end{eqnarray}
which is the Kummer equation or the confluent hypergeometric function \cite{abra}. In order to obtain a solution for the equation (\ref{3.9}) regular at the origin, we consider only the Kummer function of first kind given by $M_{s}\left(\xi\right)=M\left(\frac{\left|\gamma_{s}\right|}{2}+\frac{1}{2}-\frac{\beta_{s}}{4\Omega k},\left|\gamma_{s}\right|+1,\xi=\Omega k \rho^{2}\right)$ \cite{abra}. Therefore, a normalized radial wave function can be obtained if we impose that the hypergeometric series becomes a polynomial of degree $n$. This makes the radial wave function finite everywhere \cite{landau}. This can be achieved when the parameter $\frac{\left|\gamma_{s}\right|}{2}+\frac{1}{2}-\frac{\beta_{s}}{4\Omega k}$ is equal to a non-positive integer number, that is, when $\frac{\left|\gamma_{s}\right|}{2}+\frac{1}{2}-\frac{\beta_{s}}{4\Omega k}=-n$ (with $n=0,1,2,\ldots$). With this condition, the energy levels of the bound states are
\begin{eqnarray}
\mathcal{E}_{n,\,l}=\frac{2\Omega k}{m}\left[n+\frac{\left|\gamma_{s}\right|}{2}-\frac{\gamma_{s}}{2}+\frac{1}{2}\right]+\frac{1}{2m}\left(k+s\frac{\Omega}{2}\right)^{2}.
\label{3.10}
\end{eqnarray} 

The expression (\ref{3.10}) corresponds to the energy levels for a neutral particle with a permanent magnetic dipole moment interacting with an external electric field in an elastic Landau system. This elastic Landau system is obtained due to the presence of the topological defect, where the uniform distribution of dislocations in the elastic medium plays the role of a uniform magnetic field. 

From (\ref{3.6a}), we can see that the energy levels (\ref{3.10}) depends on the Aharonov-Casher geometric phase \cite{ac} whose periodicity is $\phi_{0}=2\pi$, that is, $\mathcal{E}_{n,\,l}\left(\phi_{\mathrm{AC}}+\phi_{0}\right)=\mathcal{E}_{n,\,l+1}\left(\phi_{\mathrm{AC}}\right)$.  Observing the harmonic-type confinement in (\ref{3.6}) due to the presence of a density of screw dislocations, we have an effective magnetic confinement analogous to a quantum dot \cite{tan2}. Thereby, we can calculate persistent spin currents \cite{ring1,ring4} in analogous way to the persistent currents observed in a quantum dot \cite{tan2}. Hence, from the dependence of the energy levels on the Aharonov-Casher geometric phase, there exists the arising of persistent spin currents in this system given by \cite{by,ring1,ring4}
\begin{eqnarray}
\mathcal{I}=-\sum_{n,\,l}\frac{\partial\mathcal{E}_{n,\,l}}{\partial\phi_{\mathrm{AC}}}=-\sum_{n,\,l}\frac{s}{2\pi}\frac{\Omega k}{m}\left[\frac{\gamma_{s}}{\left|\gamma_{s}\right|}-1\right].
\label{3.11}
\end{eqnarray}

Note that the appearance of the persistent spin currents (\ref{3.11}) in the Aharonov-Casher system comes from the presence of the uniform distribution of screw dislocation along the path of the neutral particle. Without this uniform distribution of screw dislocation, a discrete spectrum of energy would be impossible in the Aharonov-Casher system because the neutral particle would be a free particle. We also have that the presence of topological defects in this system does not change the periodicity of the persistent spin currents as pointed out in \cite{fur4} for a spinless quantum particle without torsion). The persistent spin currents (\ref{3.11}) is a periodic function of the Aharonov-Casher geometric phase $\phi_{\mathrm{AC}}$, whose periodicity is $\phi_{0}=2\pi$.

\section{hard-wall confining potential}

In this section, we consider the presence of a hard-wall confining potential in the elastic Landau system discussed in the previous section. This confining potential used here aims to describe a more realistic geometry of a quantum dot in the presence of a uniform screw dislocation density. In the following, we consider the wave function of the neutral particle being well-behaved at the origin, and vanishing at a a fixed radius $\rho_{B}$. As we have obtained in the previous section, the radial solutions (\ref{3.8}) are given by
\begin{eqnarray}
R_{s}\left(\xi\right)=e^{-\frac{\xi}{2}}\,\xi^{\frac{\left|\gamma_{s}\right|}{2}}\,M\left( \frac{\left|\gamma_{s}\right|}{2}+\frac{1}{2}-\frac{\beta_{s}}{4\Omega k},\left|\gamma_{s}\right|+1,\xi=\Omega k \rho^{2}\right).
\label{4.1}
\end{eqnarray}

In order to obtain a normalized solution inside the range $0\,<\rho\,<\,\rho_{B}$, we consider $\Omega k$ being quite small. In this way, by taking a fixed value for the parameter $b=\left|\gamma_{s}\right|+1$ of the Kummer function, we can consider the parameter of the Kummer function $a=\frac{\left|\gamma_{s}\right|}{2}+\frac{1}{2}-\frac{\beta_{s}}{4\Omega  k}$ being large due to $\Omega k$ to be small. Thus, by considering a fixed radius $\rho_{B}$, we can write the Kummer function in (\ref{4.1}) as follows \cite{abra,b10}:
\begin{eqnarray}
M\left(a,b,\xi_{0}=\Omega k\,\rho^{2}_{B}\right)&\approx&\frac{\Gamma\left(b\right)}{\sqrt{\pi}}\,e^{\frac{\xi_{0}}{2}}\left(\frac{b\xi_{0}}{2}-a\xi_{0}\right)^{\frac{1-b}{2}}\times\nonumber\\
[-2mm]\label{4.2}\\[-2mm]
&\times&\cos\left(\sqrt{2b\xi_{0}-4a\xi_{0}}-\frac{b\pi}{2}+\frac{\pi}{4}\right),\nonumber
\end{eqnarray}
where $\Gamma\left(b\right)$ is the gamma function. Next, by applying the boundary condition $R_{s}\left(\rho_{B}\right)=0$, we obtain the following expression for the energy levels:
\begin{eqnarray}
\mathcal{E}_{n,\,l}\approx\frac{1}{2m\rho_{B}^{2}}\left[n\pi+\frac{\left|\gamma_{s}\right|}{2}\pi+\frac{3\pi}{4}\right]^{2}-\frac{\Omega k}{m}\,\gamma_{s}+\frac{1}{2m}\left(k+s\frac{\Omega}{2}\right)^{2}.
\label{4.4}
\end{eqnarray}

The energy levels (\ref{4.4}) correspond to having the neutral particle interacting with a radial electric field in a elastic medium confined to a hard-wall confining potential. Despite the uniform distribution of screw dislocation plays the role of a uniform magnetic field yielding the arising of an analogue of the Landau quantization \cite{landau}, we have in this case a confinement of a neutral particle to a quantum dot with a hard-wall confining potential \cite{dot,bf20,b10}. The presence of the topological defects on this system yields two new contributions to the spectrum of energy given by the two last terms of (\ref{4.4}) in contrast to the results obtained in Ref. \cite{bf20} in the absence of defects. Moreover, we can recover the results obtained in Ref. \cite{bf20} for a neutral particle confined to a quantum dot with a hard-wall confining potential by taking the limit $\Omega\rightarrow0$. Although there exists the influence of topological defects on this system, the degeneracy of the energy levels is not broken due to the presence of defects as it has been shown in studies involving quantum rings \cite{ani}, quantum dots \cite{fur4} (without torsion), and a confinement of a neutral particle to a quantum dot induced by noninertial effects \cite{b2}. Following the discussion of Ref. \cite{ani}, the presence of torsion associated with the uniform distribution of screw dislocations introduces a spiral structure in the medium. 

Furthermore, we can see that the energy levels (\ref{4.4}) depend on the Aharonov-Casher geometric phase $\phi_{\mathrm{AC}}$ \cite{ac}, with a periodicity $\phi_{0}=2\pi$. From this dependence on the geometric phase, persistent currents arise inside the quantum dot. By using the Byers-Yang relation \cite{by}, the persistent spin currents are given by 
\begin{eqnarray}
\mathcal{I}&=&-\sum_{n,\,l}\frac{\partial\mathcal{E}_{n,\,l}}{\partial\phi_{\mathrm{AC}}}\nonumber\\
&=&-\sum_{n,\,l}\frac{s}{4m\rho_{B}^{2}} \left[n\pi+\frac{\left|\gamma_{s}\right|}{2}\pi+\frac{3\pi}{4}\right]\frac{\gamma_{s}}{\left|\gamma_{s}\right|}+\frac{s}{2\pi}\frac{\Omega k}{m}.
\label{4.5}
\end{eqnarray} 

Note that the expression of the persistent currents (\ref{4.5}) is a periodic function of the Aharonov-Casher geometric phase $\phi_{\mathrm{AC}}$, whose periodicity is $\phi_{0}=2\pi$. We can see that the presence of topological defects does not change the periodicity of the persistent currents as pointed out in Ref. \cite{fur4} in the absence of torsion. The influence of torsion on the persistent currents inside the quantum dot is given by adding a new term: $\frac{s}{2\pi}\frac{\Omega k}{m}$.  By taking the limit  $\Omega\rightarrow0$, we recover the same expression for the persistent currents obtained in Ref. \cite{bf20}.

\section{conclusions}

In this work, we have shown that the influence of a classical background made by a uniform distribution of screw dislocation on the Aharonov-Casher system \cite{ac} gives rise to bound states analogous to the Landau quantization having a dependence on the Aharonov-Casher geometric phase. In this case, the Aharonov-Casher geometric phase arises from the interaction between the permanent magnetic dipole moment of a neutral particle and an electric field produced by a charged screw dislocation placed at the center of the area $A$ that contains a uniform distribution of screw dislocations. Besides, we have seen that the uniform distribution of screw dislocation plays the role of a uniform magnetic field which gives rise to the elastic Landau system \cite{f4}. From the dependence of the energy levels on the Aharonov-Casher geometric phase, we have obtained that the energy levels are a periodic function of the geometric phase, and yields the arising of persistent currents in the elastic Landau system.

We have also discussed the confinement of a neutral particle with a permanent magnetic dipole moment to a hard-wall confining potential in the presence of an electric field. We have shown that by imposing a condition on the parameter related to the uniform distribution of defects that we can obtain bound states analogous to having a neutral particle confined to a quantum dot in the asymptotic limit. Furthermore, we have seen that the presence of topological defects does not break the degeneracy of the energy levels in contrast to previous studies of the influence of topological defects on quantum dots \cite{fur4,b2}. We also have seen that the energy levels depend on the Aharonov-Casher geometric phase, and obtained the persistent currents from this dependence on the geometric phase. Moreover, we have shown that the presence of topological defects does not change the periodicity of the persistent currents as pointed out in Ref. \cite{fur4} in the absence of torsion, but yields a new term in the expression of the persistent currents.  Note that the existence of this new contribution to the persistent current in (\ref{4.5}), which arises from the presence of torsion in a quantum dot, can be investigated in semiconductors quantum dots possessing a density of screw dislocations, since persistent spin currents can be measured in these systems.

\acknowledgments

The authors would like to thank CNPq, CAPES/NANOBIOTEC, CNPQ/PNPD, CNPQ/Universal for financial support.

\end{document}